\documentclass{PoS}

\usepackage{graphicx}

\title{Background Systematics Studies with VERITAS Data}

\ShortTitle{Background Systematics with VERITAS}

\author{\speaker{Benjamin Zitzer}\\%\thanks{A footnote may follow.}\\
        McGill University\\
        E-mail: \email{bzitzer@physics.mcgill.ca}}

\author{for the VERITAS Collaboration\\
      
        E-mail: \email{veritas@veritas.sao.arizona.edu}}

\abstract{In the analysis of VERITAS and other IACT data, it is expected to get a distribution of statistical significances in sky map bins with mean of zero and width of unity in the absence of a $\gamma$-ray signal. However, it is not uncommon to see significance distributions of width greater than unity, indicating that the background is poorly estimated and the significances in the region of interest are incorrect. This work explores the origins of these wider significance distributions and develop solutions to this issue and test these solutions on samples of VERITAS data.}

\FullConference{35th International Cosmic Ray Conference --- ICRC2017\\
		10--20 July, 2017\\
		Bexco, Busan, Korea}

\begin{document}

\section{Introduction}

VERITAS and other Imaging Air Cherenkov Telescopes (IACTs) record images of Cherenkov showers initiated by gamma rays and other types of cosmic rays. Those images are parameterized, typically assuming the images are elliptical. The parametrization of the shower is used to reconstruct the energy and sky position of the particle that initiated the air shower. However a large number of other cosmic rays are also recorded, dominated by hadron-initiated showers, which are the major source of backgrounds for the experiment. Most of the cosmic-ray showers can be removed from the analysis by using shape-based discrimination from their parameterization, but a number still remain after these gamma-ray selection cuts. IACTs are designed to detect $\gamma$ rays, which point back to their point of origin. This assumption is used to subtract the remaining background. Typically, the remaining hadron showers are subtracted by selecting another region of the sky where no gamma-ray sources are expected and scaling this region to estimate the remaining background in the region of interest (ROI) for a gamma-ray source. The significance within the ROI is calculated typically by the Li\&Ma 1983 equation 17 \cite{1983ApJ...272..317L}: 

\begin{equation}
	S = \sqrt{2}\left(N_{ON}\ln{\frac{(1+\alpha)N_{ON}}{\alpha(N_{ON}
	+N_{OFF})}} + N_{OFF}\ln{\frac{(1+\alpha)N_{OFF}}{N_{ON}
	+N_{OFF}}}\right)^{1/2}
\end{equation}

where $N_{ON}$ is the number of counts on an ON region of interest (ROI), $N_{OFF}$ is the total number of counts in one or more background (or OFF) regions, and $\alpha$ is a scaling factor between events in the ON region and the OFF regions. The generalized equation for calculating $\alpha$ is: 

\begin{equation}
	\alpha = \frac{\int Acc_{ON}(x,y,t,E)dx dy dt dE}{\int Acc_{OFF}(x,y,t,E)dx dy dt dE}
\end{equation}

where $Acc$ is the acceptance function, defined as the probability after triggering and all off line cuts of an event being reconstructed with a 
certain position and energy \cite{2007A&A...466.1219B}. The indicies $ON$ and $OFF$ refer to the acceptances in sampled on the ON and OFF region(s), respectively. This work focuses on the Ring Background Method (RBM) \cite{2007A&A...466.1219B} where the background is defined as a single annulus around the ROI, but this work should be applicable to all background estimation methods available for IACTs in the literature. For the RBM method the acceptance is typically estimated using all data not in the ROI or in any background exclusion region. Regions of the FOV with gamma-ray sources and bright stars are excluded from background and acceptance estimates. More details on how acceptance is generated for the RBM, see Section 4.

In the analysis of VERITAS and other IACT data, it is expected to get a distribution of statistical significances of sky map bins with mean of zero and width of unity in the absence of a VHE signal. However, it is not uncommon to see significance distributions of width greater than unity, indicating that the background is poorly estimated and the significances in the ROI are incorrect for sufficiently wide distributions. As a result, the criteria for a $\gamma$-ray detection (typically 5$\sigma$ above the background) can no longer be relied upon, as the chances of a false positive are increased.  

This work explores the reasons for a wider significance distributions and solutions to this issue in simulations and in VERITAS data. This work is organized as follows: the next section describes and shows a simulation that was used to recreate the problem and explore different aspects of it. The following section describes a proceedure to correct for zenith gradients in the camera, followed by a description of a `modified' Li\&Ma equation which takes into account systematic error of the background into the calculation. Finally we will show the results of validation of these techniques on VERITAS data. 

\section{Simulations}

Significance distributions calculated with the Li\&Ma 17 equation were simulated to see if the observed wider significance distributions could be recreated. These were generated with the assumption that there was no significant signal in the field of view and therefore the mean rate measured in the ROI is the same as the background region(s) when scaled by $\alpha$.  The wider significance distributions were recreated when adding in a systematic error in $\alpha$. The algorithm for the simulations are as follows:

\begin{enumerate}
\item Generate $10^{4}$ random values for $\alpha$. In this case, we assume that $\alpha$ is Gaussian distributed with a mean of 0.1 and width of 0.001. 

\item Increase $N_{ON}$ and $N_{OFF}$ each by a random amount $10^{4}$ times (once for each $\alpha$ value generated) for a time step $\Delta t$ according to: 
\begin{equation}
	N_{ON,i+1} = N_{ON,i} + Poisson(R_{bg}\Delta t) 
\end{equation}
\begin{equation}
	N_{OFF,i+1} = N_{OFF,i} + Poisson(R_{bg}\Delta t/\alpha)	
\end{equation}

where $N_{ON,0} = N_{OFF,0} = 0$ and $Poisson(x)$ is a randomly generated integer from a Poisson distribution with mean of $x$.
\item Calculate significance for each of the $10^{4}$ sets of $N_{ON}$, $N_{OFF}$, using the mean $\alpha$ value of 0.1 in Equation 2.1.

\item Fit the distribution of the significances generated in the previous steps to a Gaussian.  That width is recorded in Figure 1.

\item Repeat steps 2 through 4 for any number of time steps. The results shown in Figure 1 used a $\Delta$t of 1 hour and maximum timestep of 100 hours.
\end{enumerate}

\begin{figure}
	\centering
	\includegraphics[scale=0.37]{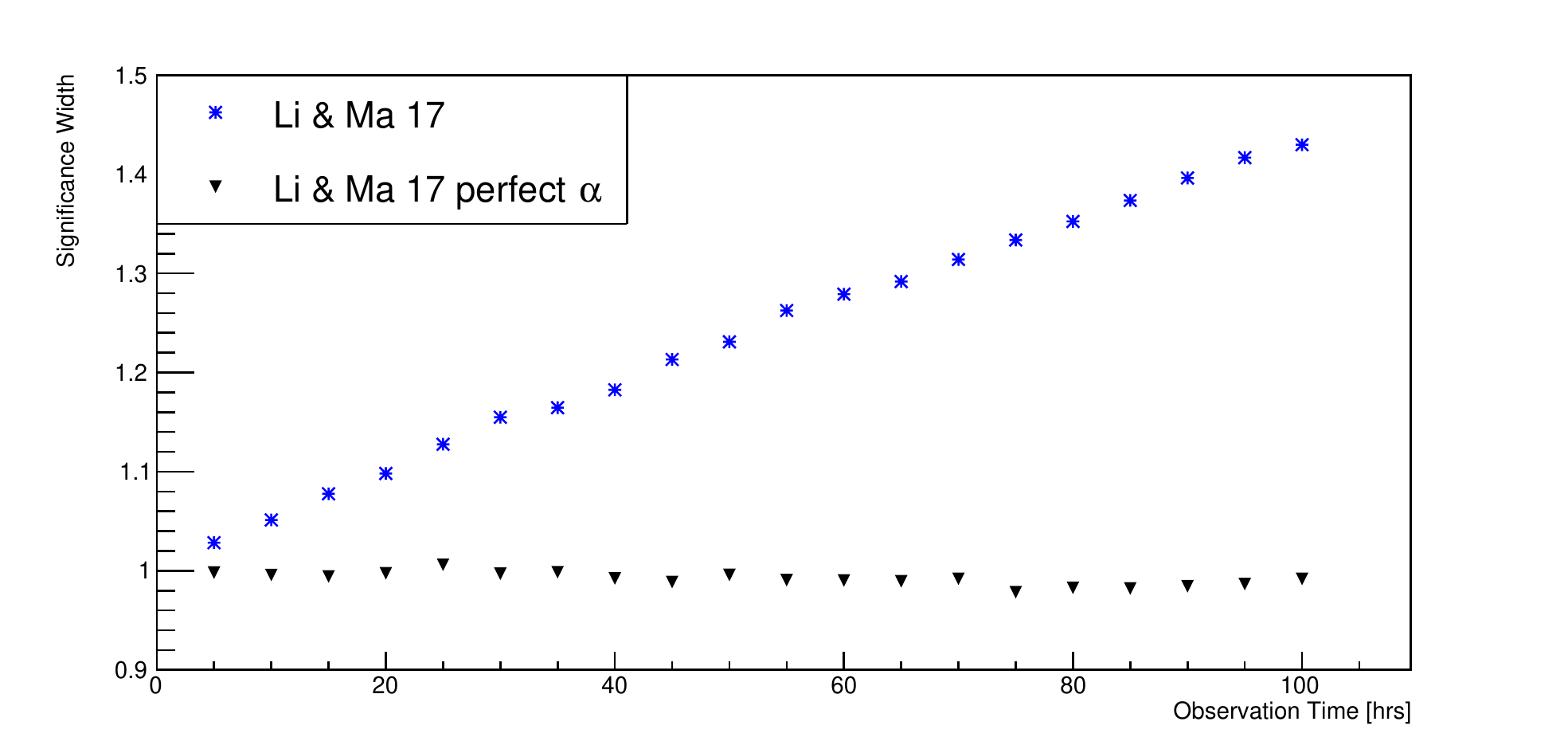}
	\includegraphics[scale=0.37]{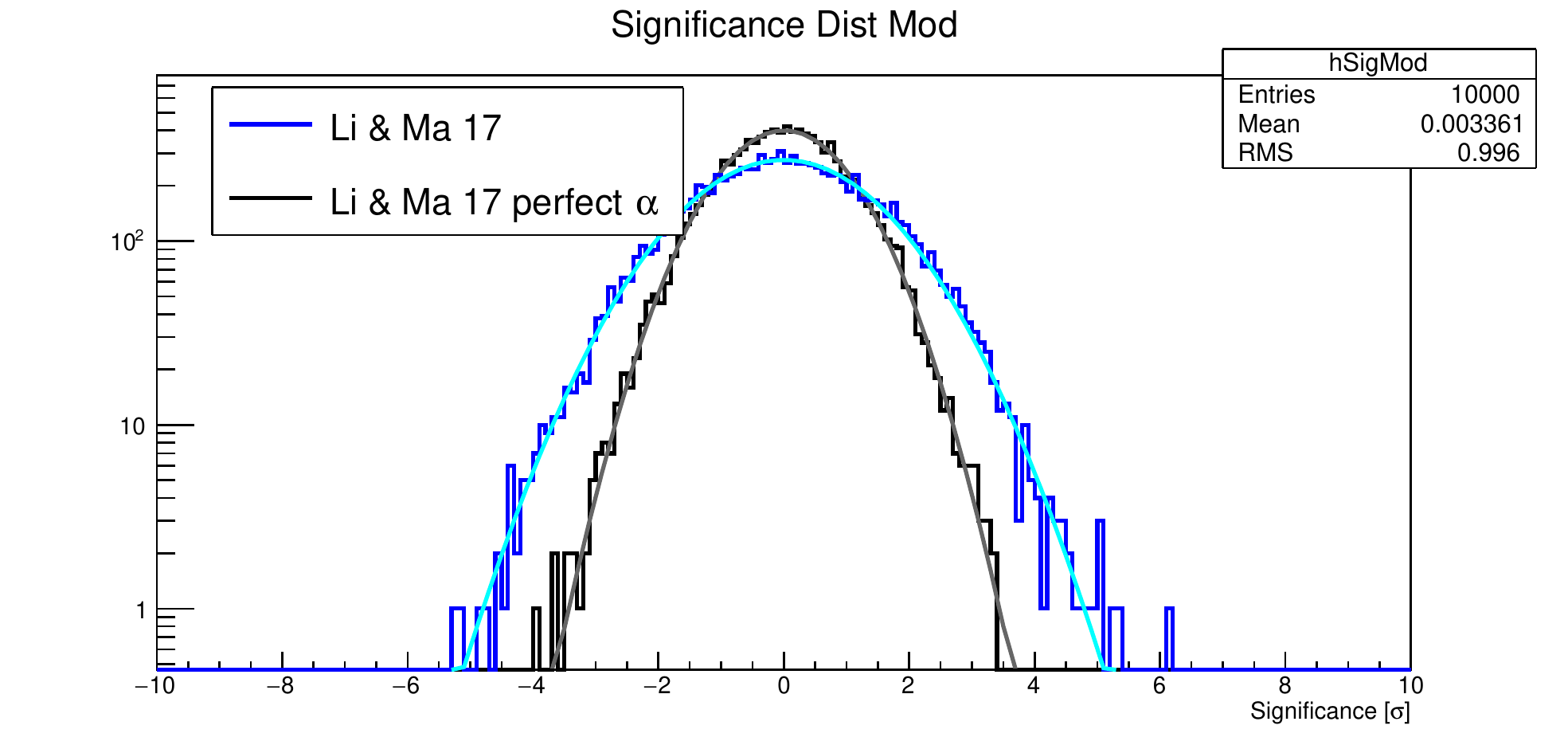}
	\caption{Example simulation results for a background rate of 2.0 cosmic rays per minute. Left: Width of the significance distribution as a function of exposure time for a `perfect' $\alpha$ and a $\alpha$ smeared by a Gaussian distribution. Right: Final significance distributions after an exposure of 100 hours. }
\end{figure}

If $\alpha$ is `perfect', i.e., the same $\alpha$ value is used in Equation 2.1 as Equation 2.4, then the width will $\it{never}$ deviate far 
from 1.0, regardless of the number of time steps. However, if we `smear' $\alpha$ by adding (or subtracting) a random small amount, $\delta\alpha$, generated from a Gaussian distribution, then the significance distribution width is $\sim$1.0 at small time steps, and eventually gets wider as 
time elapses. Figure 1 shows the results of the MC study with significance width as a function of exposure time.

Based on the simulation study, a $\delta\alpha$ value randomly generated from a Gaussian with a width $\sim0.001$, assuming a fixed $\alpha$ value of 0.1, reproduces the width of the significance distribution with the standard VERITAS analysis with similar background rates. This points to some sort of
systematic error in $\alpha$ that is not accounted for.  It is possible to tighten the $\gamma$ ray selection cuts to narrow the distribution, as this keeps the fluctuations small and the uncertainties in $\alpha$. However this option often raises the energy threshold and reduces the total statistics in the sample, so this is not the best option for many analysis, particularly for those requiring a low energy threshold. Since $\alpha$ is derived from the acceptance functions, it is reasonable to look at possible factors that could possibly affect IACT acceptance functions.

\section{Zenith Acceptance Correction}

As discussed in the previous section, issues related to the acceptance function is assumed to be responsible for the width of the significance 
distribution seen VERITAS data. The standard VERITAS analysis packages make the assumption that the acceptance function is solely a function of $\rho$, the angular distance between the tracking position of the array pointing to the reconstructed event position.  A zenith angle gradient was found in the data, which distorts the radial dependence of the acceptance function. Correcting the acceptance for the zenith gradient showed an improvement in the significance distribution width, bringing it closer to the expected distribution. This will be shown in the validation section. 

In order to qualitatively measure the acceptance gradient in the sky map, we define a new parameter called $\it{flatness}$, which is defined as the ratio of the acceptance in a given sky map bin to the number of events in that bin:

\begin{equation}
	f_{i} = \frac{N_{i}/Acc_{i}}{<N/Acc>}\sim Constant 
\end{equation}

where $f_{i}$ is the flatness, and $N_{i}$ is the number of counts in a sky map bin and the index $i$ denotes a particular sky map bin. Note that the ratio of counts to acceptance is divided by the mean of all bins so that flatness is always centered on 1.0. If the acceptance function accurately describes the background, then flatness should be approximately constant over the entire FOV for every sky map bin not in an \textit{a priori} defined background exclusion 
region. An example of a flatness map is shown in Figure 2 (left).

The next step is to produce sky maps related to zenith angle. For scaling purposes, the difference of the zenith angle of the reconstructed event position and the zenith angle of the telescope tracking position is used (Figure 2, center). The contents of each bin in the zenith map and the flatness map are plotted against each other, and the Pearson correlation coefficient is used to determine the strength of any camera gradients.  This part of the procedure was tested on Segue 1 for zenith angle, azimuth angle and NSB; only zenith angle showed any sort of strong correlation.

After the gradient was quantified for zenith angle, we then developed a proceedure to correct for it in the data. It should be noted that the both of the other major IACTs currently in operation, MAGIC \cite{Mag_Bg} and HESS \cite{HESS_Bg} use acceptance functions that are depedent in zenith, so the fact that the acceptance also takes into account zenith angle in itself is not a novel idea. The method used here makes use of the Flatness/Zenith relationship (Figure 2, right). The scatter plot of $\Delta Zn$ and flatness described in the previous section is binned into a histogram. The histogram is normalized by the value at $\Delta Zn = 0$ to better quantify the size of the effect across the camera and then is fit to a polynmial. A fourth-degree polynomial is used to remove possible second-order effects, but a linear fit most likely also acceptable. The polynomial fit is then used to re-weight the acceptance function:

\begin{equation}
	Acc'_{i} = Acc_{i}(1 + p_{1}/p_{0}\Delta Zn + p_{2}/p_{0}\Delta Zn^{2} + p_{3}/p_{0}\Delta Zn^{3} + p_{4}/p_{0}\Delta Zn^{4})
\end{equation}

Where $Acc_{i}$ is the acceptance in the $i$th RA/Dec bin, $p_{n}$ is the $n$th coefficient from the flatness fit, and $\Delta Zn$ is the 
average reconstructed zenith angle subtracted from the average telescope tracking zenith angle. Once the zenith-corrected acceptance function, 
$Acc'_{i}$ is obtained and the correction applied to each bin of the acceptance map, is used in Equation 1.2 to recalculate $\alpha$. The new $\alpha$ values are then used to recalculate significance from Equation 1.1.

\begin{figure}
	\centering
	\includegraphics[scale=0.2]{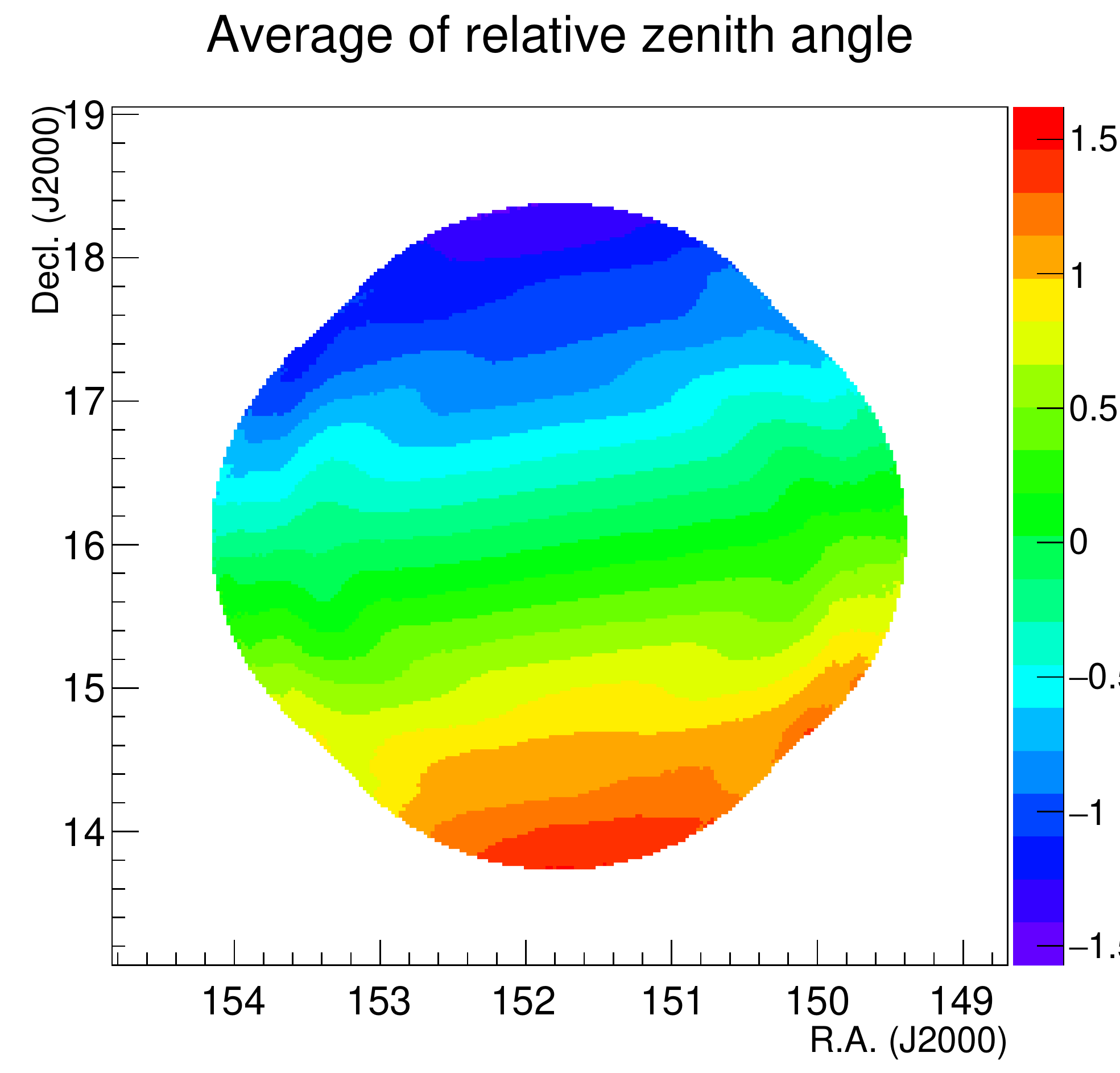}
	\includegraphics[scale=0.2]{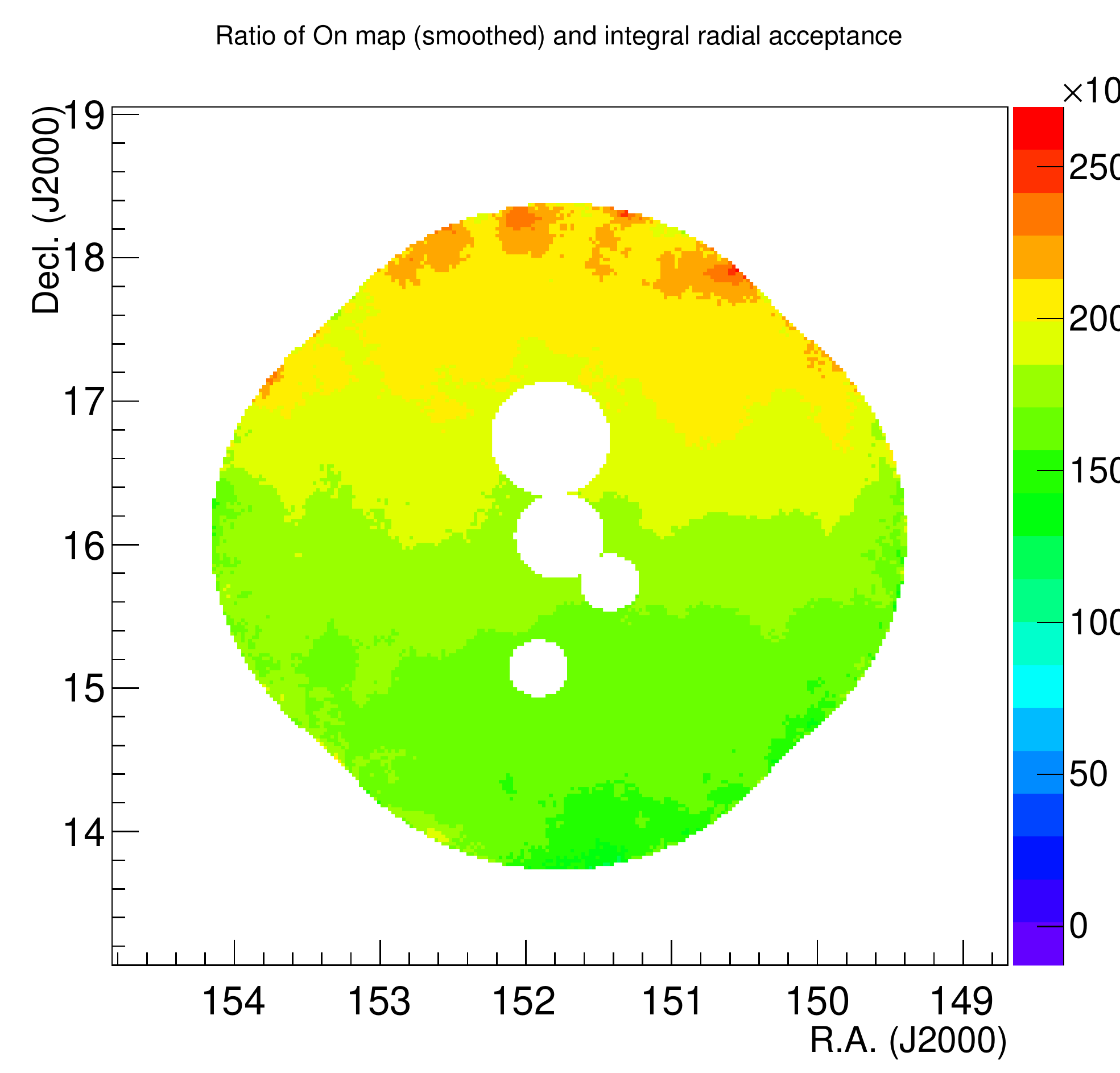}
	\includegraphics[scale=0.23]{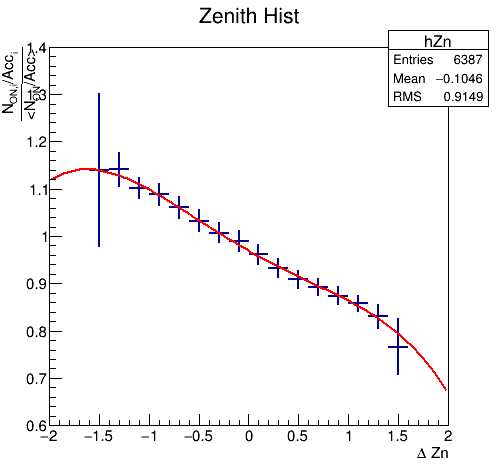}
	\caption{Visualization of the zenith gradient in VERITAS Segue 1 data. Left: Sky map of the average $\Delta$Zn, which is defined as the difference between zenith of the event reconstructed position and the zenith of the telescope tracking position. Center: Sky map showing the $N_{i}/Acc_{i}$of the VERITAS sky map, with all background exclusions regions removed. Right: Histogram of flatness as a function of $\Delta$Zn. A fit to a 4th-degree polynomial is shown in red.}
\end{figure}

\section{Modified Li\&Ma Significance Equation Accounting for Background Systematics}

Acceptance functions are typically a derived from data and there is an intrinsic systematic uncertainty associated with it. A `Modified' version of the Li\&Ma equation is needed to correctly take into account the $\alpha$ systematic error, which we will refer to as $\sigma_{\alpha} $ \cite{2015APh....67...70S}. This solution presented here is analytic, but approaches have been taken by fitting using TMinuit in the literature as well \cite{Dickinson2012}.  The full derivation of the Modified Li\&Ma Equation is in \cite{2015APh....67...70S} along with details of its performance on simulations. It follows a maximum likelihood framework with $\alpha$ treated as a nuscience parameter. The Modified Li\&Ma equation is:

\begin{equation}
	S_{Mod} = \frac{N_{on} - \alpha N_{off}}{|N_{on} - \alpha N_{off}|}\sqrt{S^{2}_{LiMa}(N_{on},N_{off},\alpha') + \Big(\frac{\alpha' - \alpha}{\sigma_{\alpha}}\Big)^{2} }
\end{equation}

Where $S_{LiMa}$ is the Li\&Ma significance calculation in Equation 1.1 and $\alpha$' is the solution to the equation:

\begin{equation}
	0 = \alpha'^{3} - \alpha'^{2}(\alpha - 1) - \alpha'(\alpha - \sigma_{\alpha}^{2}N_{off}) - \sigma_{\alpha}^{2}N_{on}
\end{equation}

This is a cubic equation with up to three real solutions. It's noteworthy to mention that in the case where $\sigma_{\alpha} = 0$, this reduces to $\alpha'$ equals $\alpha$, 0 and -1 and we recover Equation 1.1. If there are multiple positive and real solutions, then the solution is used that maximizes the likelihood:
\begin{equation}
	\Lambda(\alpha') = N_{on}\log\alpha' + (N_{on} + N_{off})\log\Big(\frac{N_{on} + N_{off}}{1+\alpha'}\Big) - \frac{1}{2}\Big(\frac{\alpha' - \alpha}{\sigma_{\alpha}}\Big)^{2}
\end{equation} 

 It is assumed here that the main source of $\sigma_{\alpha}$ is the statistical uncertainty of the acceptance curves in VEGAS, one of the VERITAS standard analysis codes \cite{VEGASRef}. The procedure for estimating $\sigma_{\alpha}$ follows, which requires how this code determines $\alpha$. On a run-by-run basis, an acceptance curve is calculated as a function of angular distance from the array pointing position to the reconstructed event position. That curve is then smoothed several times and is given a weight based on the number of events passing cuts in that run.  The weighted acceptance curve is evaluated in each sky map position in either RA/Dec or Galactic coordinates. This process is repeated for all observation runs. $\alpha$ is then calculated for each position from equation 1.2, assuming $E$ and $t$ are constant, using an ON region described as an circular region around the region of interest (ROI) and the off region being an annulus centered on the same ROI. 

The process of $\sigma_{\alpha}$ estimation follows a similar procedure as the $\alpha$ calculation. The RMS from the acceptance curve is used and the same background weight is applied. The weighted RMS is then added to each sky map position, and then the process is repeated for all runs.  $\sigma_{\alpha}$ is then calculated by error propagation of $\alpha$:

\begin{equation}
	\alpha = \frac{\int Acc_{ON}(x,y,t,E)dx dy dt dE}{\int Acc_{OFF}(x,y,t,E)dx dy dt dE} = \frac{\sum Acc_{ON}(x_{i},y_{i})}{\sum Acc_{OFF}(x_{i},y_{i})}	
\end{equation}

\begin{equation}
	\sigma_{\alpha} = \alpha\sqrt{ \Big(\frac{\sum\delta Acc_{ON}(x_{i},y_{i})}{\sum Acc_{ON}(x_{i},y_{i})} \Big)^{2} 
							+ \Big(\frac{\sum\delta Acc_{OFF}(x_{i},y_{i})}{\sum Acc_{OFF}(x_{i},y_{i})} \Big)^{2} }
\end{equation}

%\begin{figure}
%	\centering
%	\includegraphics[scale=0.22]{AcceptanceCurve.pdf}
%	\includegraphics[scale=0.22]{AcceptanceMap.pdf}
%	\includegraphics[scale=0.22]{AcceptanceErrMap.pdf}
%	\caption{From left to right: Acceptance as a function of angular distance from the center of the camera, map of acceptance in RA/Dec coordiantes, and map of the error in the acceptance.}
%\end{figure}

\section{Validation on VERITAS Data}

We test the results from these systematic studies on VERITAS data and see if they improve the significance distributions. The zenith correction and modified Li\&Ma equation were tested on various data sets. Segue 1 and PKS 1424+240 were tested using soft spectral cuts since the scientific goals of both of these benefits from having a low energy threshold; IC 443 was tested with harder but extended cuts since it is an extended SNR for VERITAS. 

Figure 3 shows the results of the validation on PKS 1424+240. Table 1 summarizes the validation results for the target position and the best fit of the significance distributions to a Gaussian function. In all cases, the significance distribution widths decrease as the zenith correction and the modified Li\&Ma are applied and the situations where both are applied the width is within 10\% of unity. In the case where both are applied, the zenith correction is applied first before determining $\sigma_{\alpha}$ for the Modified Li\&Ma equation. The zenith correction does not greatly affect the significance at the source position. Application of the modified Li\&Ma equation does decrease the target position significance, and may infer that as a loss of sensitivity. But the reader should keep in mind that if the significance distribution is sufficiently wider than unity, the significance is overestimated. Instead of a loss of sensitivity, this should be interpreted as the \textit{true} value of significance of the target observation.

\begin{table}[t]
	\begin{center}
	\scalebox{0.75}{
	\begin{tabular}{l | l | l | c | c | c | c | c | c | c }
	\hline
	Target & Zn Accept. & Mod. Li\&Ma & $N_{on}$ &  $N_{off}$ & $\alpha$ & $\sigma_{\alpha}$ & S ($\sigma$) & Sig. Mean & Sig. Width\\ 
	\hline\hline
	PKS 1424+240 & No 	& No 	& 10042 & 74779 & 0.1176 &N/A 	& 12.27  	& -0.07 & 1.43\\
	PKS 1424+240 & No 	& Yes 	& 10042 & 74779 & 0.1176 & 0.0011 	& 9.39  	& -0.07 & 1.25\\
	PKS 1424+240 & Yes 	& No 	& 10042 & 74779 & 0.1167 & N/A 	& 12.95  	& -0.08 & 1.17\\
	PKS 1424+240 & Yes 	& Yes 	& 10042 & 74779 & 0.1167 & 0.0011 	& 9.93  	& -0.07 & 1.02\\
	\hline
	Segue 1 	& No 	& No 	& 17010 & 106291 & 0.1573 &N/A 	& 2.102  & 0.02 & 1.72\\
	Segue 1  	& No 	& Yes	& 17010 & 106291 & 0.1573 & 0.0017 & 1.292  & 0.01 & 1.31\\
	Segue 1 	& Yes 	& No 	& 17010 & 106291 & 0.1590 & N/A 	& 0.782  & 0.01 & 1.23\\
	Segue 1 	& Yes 	& Yes 	& 17010 & 106291 & 0.1590 & 0.0017 & 0.479  & 0.01 & 0.99\\
	\hline
	IC443 ($\theta^{2} <$0.04)	& No	 & No   &2365 & 11199	& 0.1859	& N/A      & 5.56  & -0.03  & 1.10 \\
	IC443 ($\theta^{2} <$0.04)	& No	 & Yes  &2365 & 11199	& 0.1859	& 0.0043  & 4.00 & -0.04  & 0.95 \\
	IC443 ($\theta^{2} <$0.04)	& Yes & No   &2365 & 11199	& 0.1870	& N/A      & 5.30  &  0.08   & 1.08 \\
	IC443 ($\theta^{2} <$0.04)	& Yes & Yes  &2365 & 11199	& 0.1870	& 0.0043  & 3.81 &  0.07   & 0.93 \\
	\hline
	IC443 ($\theta^{2} <$0.09)	& No	 & No   &4913 & 11199	& 0.4040	& N/A      	& 4.79 & -0.04 & 1.16 \\
	IC443 ($\theta^{2} <$0.09)	& No	 & Yes  &4913 & 11199	& 0.4040	& 0.0093 	& 2.92 & -0.04 & 0.94 \\
	IC443 ($\theta^{2} <$0.09)	& Yes & No   &4913 & 11199	& 0.4064	& N/A     	& 4.45 & 0.09 	& 1.12 \\
	IC443 ($\theta^{2} <$0.09)	& Yes & Yes  &4913 & 11199	& 0.4064	& 0.0093 	& 2.71 & 0.08 	& 0.91 \\
	\hline
	\end{tabular}
	}
	\caption{Summary of validation results at the target position and the mean and width of the significance distributions of the validation samples. The significance distributions listed here exclude bins centered within the ROI and bright stars.}
	\end{center}
\end{table}

\begin{figure}
	\centering
	\includegraphics[scale=0.15]{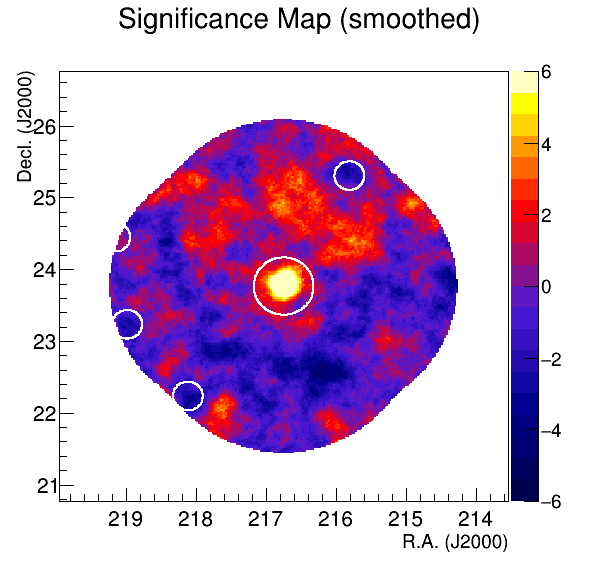}
	\includegraphics[scale=0.15]{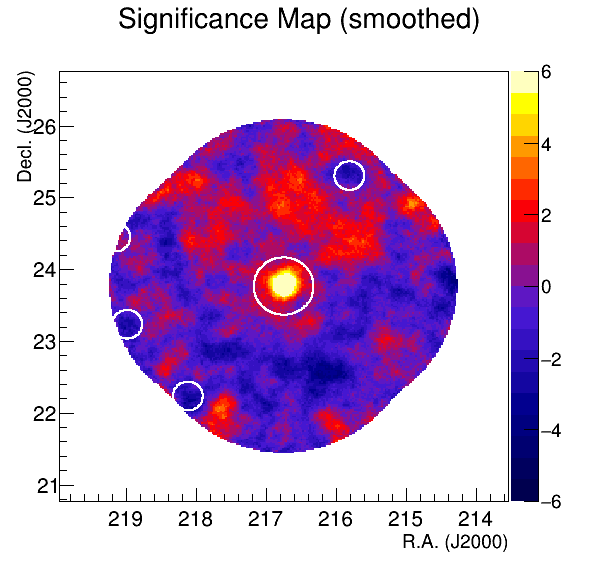}
	\includegraphics[scale=0.15]{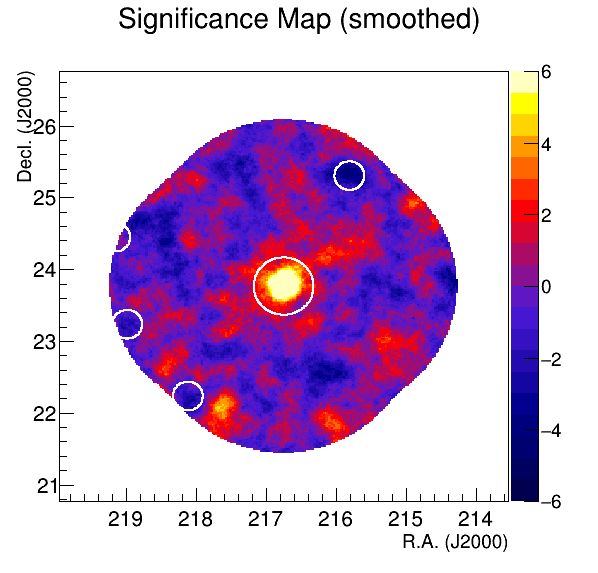}
	\includegraphics[scale=0.15]{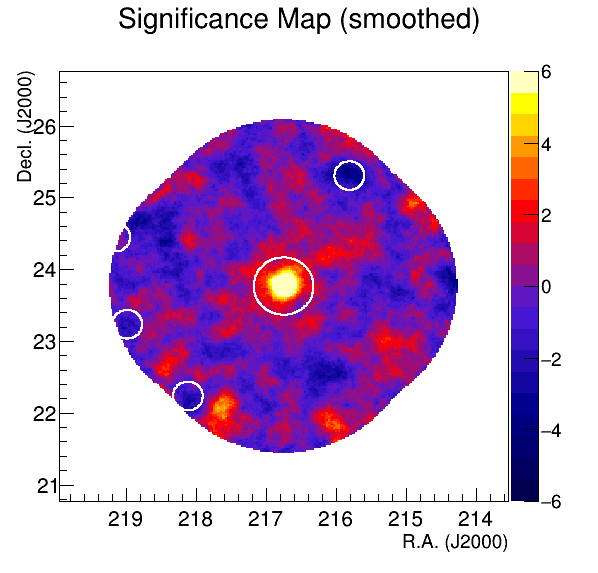}
	\includegraphics[scale=0.3]{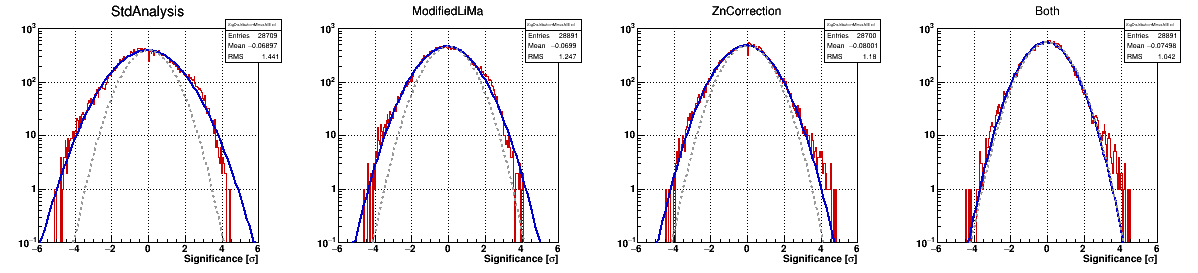}
	\caption{Top row: significance maps of PKS 1424+240. The z-axis scale is fixed between -6$\sigma$ and +6$\sigma$ in each map. Upper Far Left: significance map without zenith correction or modified Li\&Ma. Upper Center Left: significance map with modified Li\&Ma. Upper Center Left: Significance map with zenith correction. Upper Far Left: Lower Left: significance map with modified Li\&Ma and zenith correction. Bottom row: significance distributions of the field of view around PKS 1424+240 excluding the VHE source and bright stars, using the same order as the top row.}  
\end{figure}

%\begin{figure}
%	\centering
%	\includegraphics[scale=0.35]{pks1424_hFit_sigDistAll.png}
%	\caption{Significance distributions for PKS 1424+240. Distributions are shown in red, best fit to a Gaussian function is in blue, and a Gaussian with mean of zero and width of 1 is shown in gray. Left: distribution without zenith correction or modified Li\&Ma. Center left: distribution with modified Li\&Ma. Center right: distribution with zenith correction. Right: distribution with both zenith correction and modified Li\&Ma. } 
%\end{figure}

\section{Conclusions and Future Work}

This work has addressed the systematics associated with the background weighting parameter, $\alpha$, and not discussed reduction of the background rates. It is possible that other as yet unaccounted background systematics exist. Improved $\gamma$/Hadron separation would decrease the overall systematic uncertainty in the background estimation from these unaccounted for systematics and should also improve the width of significance distributions in addition to any sensitivity boost gained from a lower background rate. It is therefore encouraged to combine the methods outlined here with boosted-decision trees \cite{Krause2016} or any similar methods.

We show that employing the methods outlined here, the significance distributions in VERITAS and other IACT analysis can be greatly improved. The significance width of the validation data set of PKS 1424+240 is reduced from 1.43 to 1.02 and Segue 1 is improved from 1.72 to 0.99. Future work should investigate other factors that could affect camera acceptances, such as optically bright stars, night sky background and azimuth angle and how to best account for these effects. The techniques in this work and in the future in this area becomes incredibly important as targets observed by the current generation of IACTs for hundreds of hours or more and for the upcoming CTA observatory \cite{CTARef}. 

\section{Acknowledgments}
For the full acknowledgments, please visit https://veritas.sao.arizona.edu/.

\end{document}